# MECHANICAL EVALUATION OF SELECTIVE LASER MELTED NI-RICH NITI: COMPRESSION, TENSION, AND TORSION

**Keyvan Safaei Baghbaderani, Mohammadreza Nematollahi, Parisa Bayatimalayeri, Hediyeh Dabbaghi, Ahmadreza Jahadakbar, Mohammad Elahinia**
Dynamic and smart systems lab, The University of Toledo
Toledo, OH, USA

## ABSTRACT

*Two unique behaviors of superelasticity and shape memory effect have made shape memory alloy such as NiTi, an interesting alloy for different applications. Recently, additive manufacturing (AM) as a powerful tool for fabricating NiTi has become of interest to make complex geometries. Selective laser melting (SLM) is an AM method that not only provides flexibility to make complex 3D shapes but also it can be possible to tailor the thermomechanical properties of the parts just by changing the process parameters. The non-homogenous microstructure of as-fabricated parts as well as asymmetric mechanical behavior of NiTi, make it important to study the properties of SLM NiTi parts under different loading condition. In this study, Ni50.8Ti (at. %) powder was utilized to fabricate cube, dog-bone, and tube by SLM technique. The transformation temperatures (TTs) of samples were measured by the differential scanning calorimetry (DSC) method and the variation of TTs was discussed. Three coupons were tested mechanically under compression, tension, and torsion. In-situ digital image correlation (DIC) was employed to measure and monitor the strain field of samples during the mechanical tests. The strain distribution showed localized strain for all three samples. The equivalent stress/strain was calculated to compare the result of compressive, tensile, and torsional responses and the significant asymmetric behavior was shown and discussed.*

## 1. INTRODUCTION

Shape memory alloys (SMAs) have been of interest in many applications due to their unique properties of large recovery strain [1–4]. The large recovery strain can be achieved by inducing thermal energy, which is called shape memory effect (SME) or through the unloading which is called superelasticity. NiTi alloys are one of the most popular SMAs due to their properties such as biocompatibility, corrosion resistance, and thermomechanical stability [5]. However, because of the hard machinability of these alloys, there is a big challenge to manufacture complex shapes of NiTi alloys via conventional manufacturing techniques [6,7].

When it comes to fabricating the complex shape, additive manufacturing (AM) methods is a more preferred method. Selective laser melting (SLM) is a powder-bed fusion process that can fabricate any complex geometries just by slicing the part into finite layers [8,9]. Recently, AM methods not only are used for 3D printing of complex shapes but also they can be used as a 4D printing method to tailor the properties of the parts [10,11]. In SLM process, laser power (P), scanning speed (V), hatch spacing (H), and layer thickness (t) are the major process parameters (PPs) that tailor the micro/macro properties of the as-built samples. It's well-reported that volume energy density ($E_v = \frac{P}{VHt}$) play a key role in the final properties of the fabricated parts, while it is not the only factor [12–16]. It was reported that different PPs with the same energy density resulted in different thermomechanical behavior. Also, the microstructure can be affected significantly just by changing the one parameter a little bit [14,17]. Besides experimental and modeling studies to investigate the effect of each parameter on the microstructure and mechanical behavior of the final parts, neural network models have been employed to predict the thermomechanical properties of the SLM parts [18].

It's well-documented that the polycrystalline NiTi do not show symmetric thermomechanical behavior under different loading case [19–21]. In tension-compression loading, the NiTi has shown asymmetry mechanical behavior significantly, so that, the critical stress-induced martensite in compression mode was higher than the tensile stress [22,23]. Also, the samples under compression showed a steeper plateau with respect to tensile samples [22]. Moreover, forming the martensitic bands were reported for the NiTi samples under tension state, while no localized transformation (martensitic bands) were reported for the compression [24–26]. For the torsional loading, although no martensitic bands were reported Q. P. Sun and Z. Q. Li [27], B Reedlunn showed column localized transformation for NiTi under pure torsion [28]. The parts fabricated via SLM experience various thermal history which results in the heterogeneous microstructure. As a result, the localized transformation may have occurred due to the non-homogenous microstructure and defects formed during the process of SLM[29]. The characterization of NiTi SLM fabricated parts mostly were done on compression samples and limited data on tensile parts. While, to the best of authors' knowledge, there is no study that has reported on the torsional behavior of SLM NiTi parts.

In this study, we employed the SLM technique to fabricate dog-bone, cube, and tube NiTi coupons. The transformation temperatures (TTs) of the as-built parts were reported and the variation of the TTs was discussed. The as-fabricated samples were tested under tensile, compressional, and torsional loading. The in-situ digital image correlation (DIC) method was utilized to capture the mechanical behavior of the parts under mechanical loading. Also, the strain distribution and asymmetric behavior of the SLMed NiTi were discussed for various stress states.

## 2. MATERIALS AND METHODS

Ni50.8Ti (at. %) ingot was gas atomized to produce NiTi powder and then sieved with 25-75 microns meshes. A 3DSystems ProX200 SLM machine was used for the fabrication of samples. Compression, tension, and torsion samples were fabricated to show the behavior of SLM NiTi under different loading conditions. Compression, tension, and torsion samples were in the shape of cuboid, dog-bone, and tube, respectively with the dimensions specified in Figure 1.

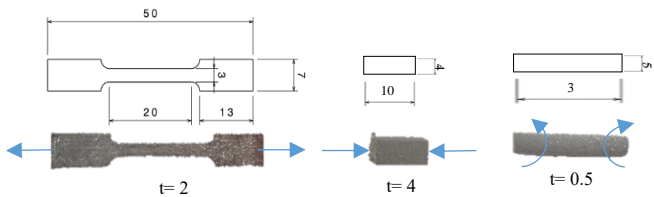

**Figure 1**. SLM fabricated coupons and dimensions

All the specimens were fabricated using the same process parameters as well as the same scanning strategy. The laser power of 250 W, scanning speed of 1250 mm/sec, hatch spacing of 80 um, and layer thickness of 30 um were chosen based on the previous studies [13,17] which result in dense parts with good superelasticity behavior. The resultant energy density was 83.33 J/mm$^3$. Tensile and compression specimens were tested using a Testresources hydraulic mechanical testing machine with a strain rate of 0.0002 s$^{-1}$ up to approximately 3% of strain and then unloaded to zero loads. An electro force multiaxial Bose machine was used for the torsion with an angle rate of 0.08 deg/s. In both cases, samples were specked with black and white spray paints and strains were measured using a Digital Image Correlation (DIC) system manufactured by Correlated Solutions. Strains were calculated using the VIC-3D software supplied by the manufacturer. All mechanical tests were performed at room temperature (28 °C) To find transformation temperatures, 25-45 mg samples were cut from the SLM fabricated specimens and Differential Scanning Calorimetry (DSC) was performed using a TA instrument DSC with a heating/cooling rate of 10 deg/min in a Nitrogen atmosphere.

## 3. RESULTS AND DISCUSSION

Figure 2 shows the DSC results of the SLM parts for three different shapes. The transformation temperatures (TTs) were reported in Table1. Each sample was cycled twice to make sure of stable transformation. The general trend shows no significant change to a slight decrease in TTs after fabrication for all samples. The high temperature of the SLM process caused impurities to be picked up during the melting. Ti is a reactive element that can make a reaction with impurities such as oxygen at the elevated temperatures. As a result, the main matrix depletes from Ti and the TTs shifts down. As it's shown, there is a slight variation between the TTs of different coupons, while the process parameters and the powder are the same for all samples. The thermal history affected the parts during the laser melting process was different for various shapes which can cause different microstructures. The various microstructure can explain the slight variation in TTs from one shape to another one. But, it needs more investigation to explain this observation.

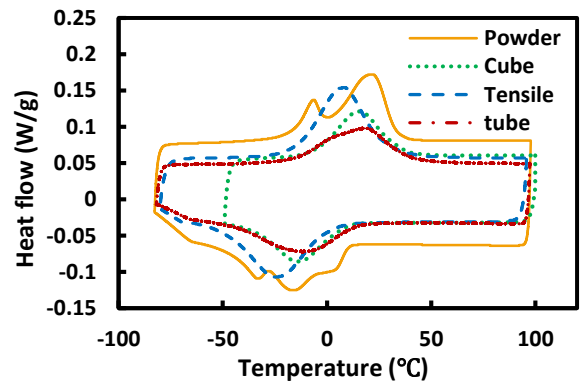

**Figure 2.** DSC result of SLM NiTi coupons

**Table 1.** The transformation temperatures of SLMed NiTi parts

| TTs* \ Sample | Ms (°C) | Mf (°C) | As (°C) | Af (°C) |
|---|---|---|---|---|
| Powder | 10.2 | -46.6 | -16.9 | 33.1 |
| Dog-bone | 2.1 | -60.1 | -15.1 | 27 |
| Tube | 16.4 | -68.7 | -13.5 | 41.9 |
| Cube | 7.4 | -32.2 | -6.2 | 35.3 |

* Ms and As are martensite and austenite start and Mf and Af are martensite and austenite finish temperatures, respectively.

To see the mechanical behavior, samples tested under pure tension, compression and torsion mode. The behavior of the cube sample under compressive loading was shown in Figure 3. The strain distribution during loading/unloading was presented for several points. The average strain was determined to plot the stress-strain curve. The sample was loaded up to 3.2% of compressive strain and unloaded to no-load condition. The as-fabricated material could recover more than 2.6% strain (78% of total strain) and had around 0.6 % of irrecoverable strain that can be the result of permanent dislocations or non-transformed detwinned martensite. The DIC images show that the strain is not homogenous and some localized strain can be detected. However, no martensite band was found which is consistent with the literature. The localized transformation can be explained by non-homogeneity of the SLM parts as discussed earlier.

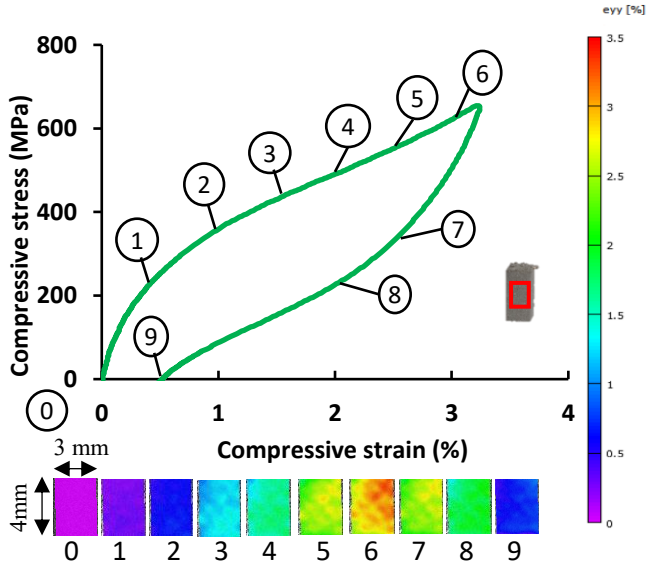

**Figure 3.** The mechanical behavior of the SLM NiTi cube sample under compression. The strain fields were presented for ten points for a complete cycle. The red box on the sample shows the DIC field of view.

Figure 4 shows the stress-strain response of the dog-bone sample under tensile loading. The sample was loaded to around 3% of strain and then unloaded to zero load state. The transformation strain of 2.3% (76% of total strain) was achieved by the as-fabricated sample under tension. Under tension, the high localized strain was formed by reaching the stress to the plateau level and then this localized strain propagated to the other regions of the material. For the NiTi manufactured by conventional methods, the formation of macroscopic austenite to martensite transformation fronts (martensitic bands) under tension has been reported in several studies, while, it's not reported yet for SLM parts. Therefore, further investigation is needed to study the strain distribution of SLM fabricated parts under tension.

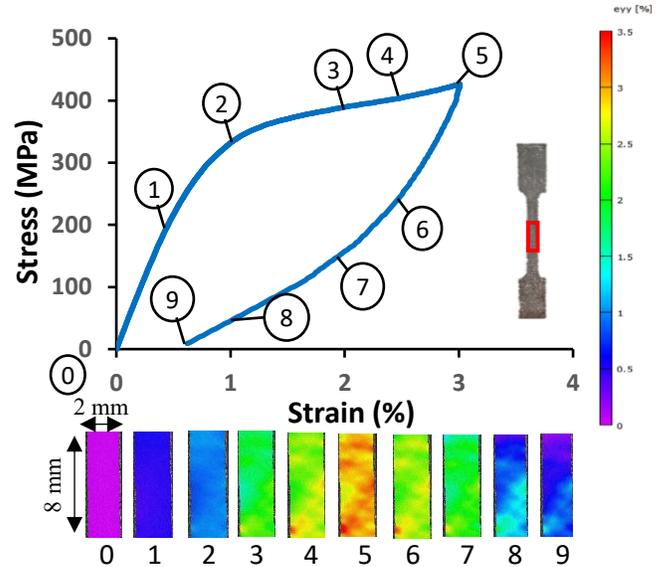

**Figure 4.** The mechanical behavior of the SLM NiTi dog-bone sample under tension. The strain fields were presented for ten points for a complete cycle. The red box on the sample shows the DIC field of view.

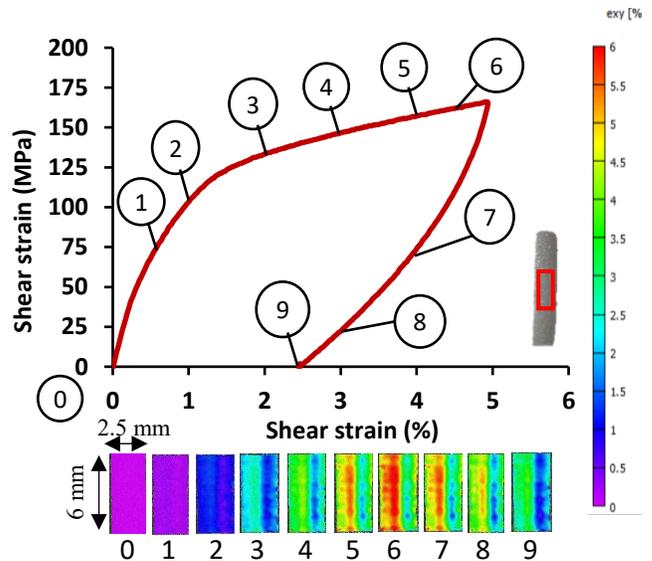

**Figure 5.** The mechanical behavior of the SLM NiTi tube sample under torsion. The strain fields were presented for ten points for a complete cycle. The red box on the sample shows the DIC field of view.

To capture the torsional behavior of the SLM fabricated coupons, the as-fabricated tube was loaded/unloaded under pure torsional loading with the low angle rate to make sure of isothermal condition. The shear stress was calculated by Equation (1) and the shear strain was measured from the 3D DIC strain field. $T(N.m), \rho(mm), and\ J(mm^4)$ are torque, the outer radius of the tube, and polar moment of inertia, respectively.

$$\tau = \frac{T.\rho}{J} \qquad (1)$$

The tube was loaded under torsional loading up to 5% of shear strain and unloaded to zero torque (Figure 5). The tube recovered about 2.5% of shear strain (50% of total strain). The large residual strain was expected to the TTs of the SLM tube which resulted in mixed-phase of austenite and martensite at room temperature. This residual strain can be recovered upon heating the sample to elevated temperature. The DIC strain field shows vertical columns of localized shear strain which have a higher strain with respect to other regions. A similar localized strain column was reported for the cold-drawn slightly Ni-rich NiTi tube by Reedlunn et al. [28]. They studied the behavior of the NiTi tube under the proportional strain path of tension and torsion. Under pure torsional loading, the columns of high shear strain were detected by the DIC method; however, no martensitic band or localized shear strain was captured by Sun and Li [27]. It's noteworthy to mention that Sun and Li used optical microscopy to detect any high localized strain and they did not have a strain field, collected through DIC technique. As it's discussed, because of having limited data on torsional behavior and strain distribution of SLM fabricated NiTi tubes, more investigations are needed to be done, to confirm and explain the observed phenomena. Also, it's good to mention that the strain filed got from DIC measurement is highly impacted by test condition, any misalignment can result in a wrong data. To avoid the test setup error, a conventional metal (e.g. Aluminum) with a same geometry should be tested under the same test condition [28].

To have a better overview of SLM NiTi behavior under different loading case, all stress-strain curve was shown in Fig.5. In Equation (2), the equivalent stress/strain based on Von-Mises criteria was measured for comparison between compression, tension, and torsion tests. $\sigma(\varepsilon)\ and\ \tau(\gamma)$ are compressive/tensile stress (strain) and shear stress(strain) respectively.

$$\sigma_e = \sqrt{\sigma^2 + 3\tau^2}$$
$$\varepsilon_e = \sqrt{\varepsilon^2 + \frac{\gamma^2}{3}} \qquad (2)$$

The equivalent stress/strain behaviors of as-fabricated parts under compression, tension, and torsion were shown in Figure 6. The large asymmetric behavior of SLM NiTi parts in uniaxial tension/compression can be seen in Fig. 5. It's shown clearly that the critical stress-induced martensite for compression is higher than tension. Moreover, the transformation plateau is steeper in compression in comparison to the tensile sample. These two observations have been also reported for conventional NiTi parts. In torsional loading, the critical stress and the slope of the transformation plateau is much lower than tension and compression samples. In addition, the shear moduli of elasticity (G) for pure torsional behavior is around 31 GPa for the as-fabricated tube, while the value which is calculated from elastic theory $G = \frac{E}{2(1+\vartheta)}$ ($E_{tensile} = 51 MPa, \vartheta_{tensile} = 0.4$) is 18.2 GPa. The Poisson ratio ($\vartheta$) of the tensile sample was calculated from the ratio of the transverse strain to the axial strain from DIC measurement.

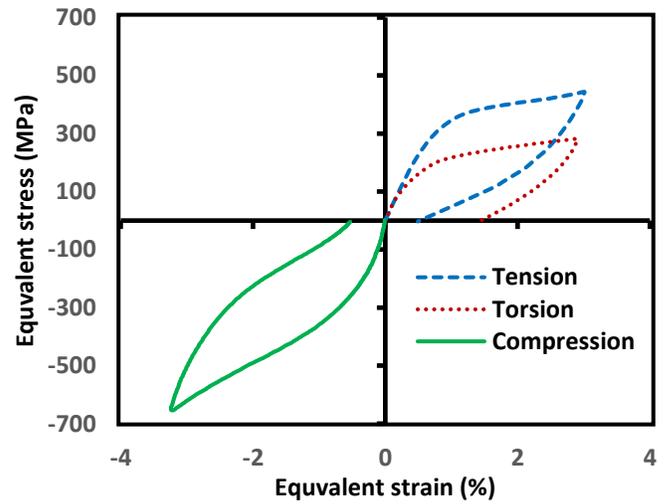

**Figure 6.** The equivalent stress-strain response of SLM NiTi parts under compressive, tensile, and torsional loading

The asymmetric behavior presented here shows the importance of a comprehensive study on SLM fabricated NiTi parts under different stress states. The actuators and parts made of NiTi mostly experience complex loading conditions, and it's not possible to design and model them without the comprehensive knowledge of their thermomechanical behavior under various conditions. Two factors can result in the asymmetric behavior of SLM NiTi parts: i) the intrinsic asymmetric behavior of NiTi alloy and ii) the non-homogenous microstructure of as-fabricated SLM parts. It's well documented that the crystallography direction and the grain size impact the thermomechanical behavior of NiTi alloys significantly [30–34]. The preferred direction of cubic B2 crystals of austenite phase in solidification is [001] [35,36]. Moreover, the heat gradient of the SLM process forms columnar grains which grow from the substrate to the top surface. As a result, the SLM as-fabricated parts has the columnar grains and strong texture in [001] direction [17]. Therefore, when it comes to additive manufacturing processes such as SLM, the effect of the microstructure is of importance and should be considered to make a comprehensive comparison between different conditions.

## 4. CONCLUSION

The SLM technique was employed to fabricate three different coupons of compression, tension, and torsion. The TTs of the parts shifted to higher temperatures with respect to the powder due to the Ni evaporation during the SLM process. Three coupons were tested mechanically, under compressive, tensile, and torsional loading. The strain field captured from DIC measurement showed localized strain for the cube sample. The localized columnar shear strain was shown for the tube sample under torsion. Under tension, the high local strain was formed at the start of the plateau and propagated to the other regions by increasing the strain. The equivalent stress/strain responses of all samples were compared to each other. It was shown significant tension/compression asymmetric behavior for SLM NiTi parts. The critical stress-induced martensite, the slope of the plateau, and the modulus of elasticity were remarkably different for various loading scenarios.

## REFERENCES

[1] M. Nematollahi, K.S. Baghbaderani, A. Amerinatanzi, H. Zamanian, M. Elahinia, Bioengineering 6 (2019) 37.
[2] H. Nuruddin, I.M. Kamal, M.N. Mansor, N.M. Hafid, ARPN J. Eng. Appl. Sci. 12 (2017) 1254–1259.
[3] M. Nematollahi, R. Mehrabi, M.A. Callejas, H. Elahinia, M. Elahinia, in: Act. Passiv. Smart Struct. Integr. Syst. XII, International Society for Optics and Photonics, 2018, p. 105952J.
[4] A. Jahadakbar, M. Nematollahi, K. Safaei, P. Bayati, G. Giri, H. Dabbaghi, D. Dean, M. Elahinia, Metals (Basel). 10 (2020) 151.
[5] M. Elahinia, Shape Memory Alloy Actuators Design , Fabrication , And, 2016.
[6] M. Elahinia, N. Shayesteh Moghaddam, M. Taheri Andani, A. Amerinatanzi, B.A. Bimber, R.F. Hamilton, Prog. Mater. Sci. 83 (2016) 630–663.
[7] M. Nematollahi, A. Jahadakbar, M.J. Mahtabi, M. Elahinia, in: Met. Biomed. Devices (Second Ed., Woodhead Publishing Series in Biomaterials, 2019, pp. 331–353.
[8] S.E. Saghaian, N.S. Moghaddam, M. Nematollahi, S. Saedi, M. Elahinia, H.E. Karaca, Eng. Med 3 (2018) 1–7.
[9] A. Jahadakbar, N.S. Moghaddam, A. Amerinatanzi, D. Dean, M. Elahinia, in: Behav. Mech. Multifunct. Mater. Compos. XII, International Society for Optics and Photonics, 2018, p. 1059610.
[10] J. Ma, B. Franco, G. Tapia, K. Karayagiz, L. Johnson, J. Liu, R. Arroyave, I. Karaman, A. Elwany, Sci. Rep. 7 (2017) 1–8.
[11] T. Bormann, R. Schumacher, B. Müller, M. Mertmann, M. de Wild, J. Mater. Eng. Perform. 21 (2012) 2519–2524.
[12] M. Nematollahi, G. Toker, S.E. Saghaian, J. Salazar, M. Mahtabi, O. Benafan, H. Karaca, M. Elahinia, Shape Mem. Superelasticity 5 (2019) 113–124.
[13] S. Saedi, N. Shayesteh Moghaddam, A. Amerinatanzi, M. Elahinia, H.E. Karaca, Acta Mater. 144 (2018) 552–560.
[14] N. Shayesteh Moghaddam, S.E. Saghaian, A. Amerinatanzi, H. Ibrahim, P. Li, G.P. Toker, H.E. Karaca, M. Elahinia, Mater. Sci. Eng. A 724 (2018) 220–230.
[15] G.P. Toker, M. Nematollahi, S.E. Saghaian, K.S. Baghbaderani, O. Benafan, M. Elahinia, H.E. Karaca, Scr. Mater. 178 (2020) 361–365.
[16] C.A. Biffi, P. Bassani, M. Nematollahi, N. Shayesteh Moghaddam, A. Amerinatanzi, M.J. Mahtabi, M. Elahinia, A. Tuissi, Materials (Basel). 12 (2019) 3068.
[17] N. Shayesteh Moghaddam, S. Saedi, A. Amerinatanzi, A. Hinojos, A. Ramazani, J. Kundin, M.J. Mills, H. Karaca, M. Elahinia, Sci. Rep. 9 (2019) 1–11.
[18] M. Mehrpouya, A. Gisario, A. Rahimzadeh, M. Nematollahi, K.S. Baghbaderani, M. Elahinia, Int. J. Adv. Manuf. Technol. (2019) 1–9.
[19] P. Sittner, Y. Hara, M. Tokuda, Metall. Mater. Trans. A 26 (1995) 2923–2935.
[20] D.J. Hartl, A. Solomou, D.C. Lagoudas, D. Saravanos, Behav. Mech. Multifunct. Mater. Compos. 2012 8342 (2012) 83421M.
[21] T.J. Lim, D.L. Mcdowell, (1999) 1–2.
[22] S.C. Mao, J.F. Luo, Z. Zhang, M.H. Wu, Y. Liu, X.D. Han, Acta Mater. 58 (2010) 3357–3366.
[23] D. Favier, L. Orgéas, Acta Mater. 46 (1998) 5579–5591.
[24] B. Reedlunn, C.B. Churchill, E.E. Nelson, J.A. Shaw, S.H. Daly, J. Mech. Phys. Solids 63 (2014) 506–537.
[25] J.A. Shaw, S. Kyriakides, Acta Mater. 45 (1997) 683–700.
[26] Z.Q. Li, Q.P. Sun, Int. J. Plast. 18 (2002) 1481–1498.
[27] Q.P. Sun, Z.Q. Li, Int. J. Solids Struct. 39 (2002) 3797–3809.
[28] B. Reedlunn, S. Daly, J. Shaw, in: Proc. ASME 2012 Conf. Smart Mater. Adapt. Struct. Intell. Syst. SMASIS2012, 2016, pp. 1–10.
[29] B.A. Bimber, R.F. Hamilton, J. Keist, T.A. Palmer, Mater. Sci. Eng. A 674 (2016) 125–134.
[30] K. Gall, H. Sehitoglu, Int. J. Plast. 15 (1999) 69–92.
[31] C. Qiu, N.J.E. Adkins, M.M. Attallah, Mater. Sci. Eng. A 578 (2013) 230–239.
[32] S. Saedi, A.S. Turabi, M.T. Andani, N.S. Moghaddam, M. Elahinia, H.E. Karaca, Mater. Sci. Eng. A 686 (2017) 1–10.
[33] I. Kaya, H.E. Karaca, M. Souri, Y. Chumlyakov, H. Kurkcu, Mater. Sci. Eng. A 686 (2017) 73–81.
[34] J.-L. Liu, H.-Y. Huang, J.-X. Xie, Mater. Des. 85 (2015) 211–220.
[35] K. Gall, M.L. Dunn, Y. Liu, P. Labossiere, H. Sehitoglu,

Y.I. Chumlyakov, J. Eng. Mater. Technol. 124 (2002) 238–245.

[36] H. Sehitoglu, I. Karaman, R. Anderson, X. Zhang, K. Gall, H.J. Maier, Y. Chumlyakov, Acta Mater. 48 (2000) 3311–3326.